\documentclass[utf8]{FrontiersinHarvard} 

\usepackage{url,hyperref,lineno,microtype,subcaption}
\usepackage[onehalfspacing]{setspace}

\usepackage{booktabs}
\usepackage{xcolor}
\usepackage[normalem]{ulem} 
\definecolor{darkorange}{RGB}{180,100,00}
\definecolor{darkblue}{RGB}{0,0,139}

\def\keyFont{\fontsize{8}{11}\helveticabold }
\def\firstAuthorLast{Delbo {et~al.}} 
\def\Authors{Marco Delbo\,$^{1,2,*}$, 
	Thomas Dyer\,$^{1}$,  
	Ullas Bhat\,$^{2}$,  
	Chrysa Avdellidou\,$^{2}$,
	Laurent Galluccio\,$^{1}$,  
	Amelia Milton\,$^{2,3}$}
\def\Address{$^{1}$Université Côte d'Azur, CNRS-Lagrange, 
Observatoire de la Côte d'Azur, CS 34229 - F 06304 NICE Cedex 4, France \\
$^{2}$University of Leicester, School of Physics and Astronomy, University Road, LE1 7RH, Leicester, UK \\
$^{3}$University of Leicester, School of Natural Sciences, University Road, 
LE1 7RH, Leicester, UK}
\def\corrAuthor{Marco Delbo} 
\def\corrEmail{delbo@oca.eu}

\begin{document}
\onecolumn
\firstpage{1}

\title[Gaia DR3 asteroid spectral classification]{Gaia DR3 supervised classification of asteroid reflectance spectra}

\author[\firstAuthorLast ]{\Authors} 
\address{} 
\correspondance{} 

\extraAuth{}

\maketitle

\begin{abstract}

\section{}
We present a supervised, probabilistic taxonomic classification of asteroid reflectance spectra from Gaia Data Release 3 (DR3). Using high-quality Gaia DR3 spectra and a reference set of spectra from the literature consisting exclusively of asteroids with robust spectroscopic taxonomic types, we construct a principal-component (PC) representation of the Gaia reflectances. For each major spectral complex (C, S, X) and several end-member classes (B, D, A, L, K, V), we model the distribution of reference objects in PC space using multivariate kernel density estimation (KDE). This yields  likelihoods for each spectral class and provides a quantitative measure of classification confidence. 

Validation against a sample of objects with known spectral classes demonstrates good performance for classes with distinctive reflectance signatures, including the S-complex, D, V, and A types. Spectrally continuous classes (B-C-complex, K-L-S-complex, and X-complex) show the expected degrees of mixing given the limited wavelength range of Gaia's spectrophotometry. We further explore the compositional structure of six major asteroid collisional families using our Gaia-derived spectral classes, finding excellent agreement with ground-based spectroscopy and revealing enhanced detections of olivine-rich A type material in the Flora and Eunomia families, as well as new insights into the spectral diversity of the Tirela family.

The resulting catalogue constitutes a fully probabilistic taxonomic classification for the full Gaia DR3 asteroid sample. It offers a resource for studying the compositional structure of the main belt, identifying family interlopers, and linking asteroid populations to meteorite groups, and establishes a methodological framework for future Gaia releases, in particular for the validation of the Gaia DR4, expected by the end of 2026.

\tiny
 \keyFont{ \section{Keywords:} Asteroids, Asteroid physical properties, Gaia, Spectroscopy, Spectral classification, Astronomical-data} 
\end{abstract}

\section{Introduction}

Planetesimals are the first objects that formed in the protoplanetary disk of our solar system within the first Myr of its evolution \citep{klahrFormationMainBelt2022}. Planetesimals accreted relatively big, with diameters $\gtrsim$30 km \citep{morbidelliAsteroidsWereBorn2009,delboIdentificationPrimordialAsteroid2017,delboAncientPrimordialCollisional2019,ferroneIdentification43Billion2023}, at different heliocentric distances that led to distinct compositions \citep{kruijerGreatIsotopicDichotomy2019,andersonDifferentArrivalTimes2025}. Their earliest generation underwent differentiation due to the presence of radioactive elements \citep{weissDifferentiatedPlanetesimalsParent2013}, while undifferentiated ones that kept a chondritic composition, formed a few Myr later. Dynamical processes \citep{tsiganisOriginOrbitalArchitecture2005,walshLowMassMars2011,raymondEmptyPrimordialAsteroid2017} operating during the first 100 Myr of Solar System evolution \citep{nesvornyTransNeptunianBinariesEvidence2019,avdellidouDatingSolarSystems2024} may have placed remnants of the planetesimal population in locations distinct from their original accretion regions \citep{levisonContaminationAsteroidBelt2009,vokrouhlickyCaptureTransneptunianPlanetesimals2016,avdellidouDatingSolarSystems2024}. The main belt between Mars and Jupiter is an example of such material mixing, as it may host objects that originated in multiple, distinct regions of the early Solar System \citep{demeoCompositionalStructureAsteroid2015}.

However, only a small fraction of the original planetesimal population has survived, as the majority were either incorporated into growing planets, ejected from the Solar System, or disrupted by large collisions that produced the asteroid collisional families observed today \citep{nesvornyIdentificationFamiliesAstIV2015,delboIdentificationPrimordialAsteroid2017,dermottCommonOriginFamily2018}. The current asteroid populations preserve in their mineralogical and physical properties a record of the processes that occurred in the early Solar System, including accretion, differentiation, and collisional evolution. The study of asteroids therefore provides a unique window into the conditions and mechanisms that led to the formation of the planets.

Asteroids are the parent bodies of most of the meteorites that currently exist in our collections. Therefore, establishing a link between the meteorite groups and their source asteroids remains a key challenge in planetary science with important implications for understanding planetesimal composition, accretion time and place, thermal history, dynamical and collisional evolution \citep{henkeThermalHistoryModelling2012,trieloffEvolutionParentBody2022,avdellidouAthorAsteroidFamily2022a,avdellidouDatingSolarSystems2024,brozSourceRegionsCarbonaceous2024,brozYoungAsteroidFamilies2024,marssetMassaliaAsteroidFamily2024}. In order to establish asteroid-meteorite links several physical properties need to be studied and matched, with reflectance spectra being one of the most crucial.

Over the past decades, several ground-based telescopes provided reflectance spectra for more than 6,500 asteroids and collected spectrophotometric data for more than 370,000 asteroids as presented in Minor Planet Physical Properties Catalogue \citep[MP3C;][\url{https://mp3c.oca.eu}; accessed Dec 4th, 2025]{delboMinorPlanetPhysical2022,dyerDependenceAsteroidRotation2026}. Based on these data, several taxonomic systems have been developed to classify asteroids according to their spectra in visible and near-infrared wavelengths \citep{tholenASTEROIDTAXONOMYCLUSTER1984,busPhaseIISmall2002,demeoExtensionBusAsteroid2009,mahlkeAsteroidTaxonomyCluster2022}, as well as according to their broadband photometry from large-scale surveys such as the Moving Object Catalog of the Sloan Digital Sky Survey (MOC SDSS) and similar surveys \citep{demeoTaxonomicDistributionAsteroids2013,carvanoSDSSbasedTaxonomicClassification2010,sergeyevMillionAsteroidObservations2021,sergeyevMultifilterPhotometrySolar2022,choiTaxonomicClassificationAsteroids2023}. Spectral classes are  regarded as a proxy for surface composition, condensing an object's reflectance properties and key spectral features into a single descriptive parameter: the taxonomic class.

All available taxonomic classes were compiled into a database by combining classifications from multiple spectroscopic and photometric surveys \citep{dyerDependenceAsteroidRotation2026}: For each asteroid, \citet{dyerDependenceAsteroidRotation2026} collected all published class assignments together with their associated quality indicators and observational metadata. A “best class” was determined by selecting the most reliable classification based on signal-to-noise ratio, consistency across surveys, and internal scoring criteria. The resulting catalogue provides, for each object, both the full set of reported classifications and a single recommended taxonomic type used in further analyses. This catalogue is publicly available through MP3C.

The advent of the Gaia mission of the European Space Agency (ESA) has opened a new era in asteroid spectrophotometry. Gaia Data Release 3 (DR3) includes low-resolution reflectance spectra for 60,518 small bodies \citep{gaiacollaborationGaiaDataRelease2023}, representing a large and homogeneous dataset of asteroid spectrophotometry. Despite its limited spectral resolution and some biases, as detailed later, the Gaia dataset offers a unique opportunity to derive physical and compositional properties for an extremely large sample, extending taxonomic classification far beyond the reach of ground-based spectroscopy.

Previous studies have applied various methods to classify Gaia DR3 asteroid spectra. \citet{muinonenDR3classification2023} presented a neural network-based classification method combining Gaia spectroscopy and photometric slopes. \citet{geAsteroidTypesAlbedos2025} applied the neural network framework of \citet{geAsteroidMaterialClassification2024}, trained on a combination of reflectance spectra, albedo information, and orbital elements, to classify asteroids into taxonomic types and to infer their albedos and diameters using their absolute magnitudes.

One of the most classical tools to identify the major compositional classes from the reflectance spectra is historically the use of the principal component analysis (PCA) 
\citep{tholenASTEROIDTAXONOMYCLUSTER1984,busPhaseIISmall2002,roigSelectingCandidateVtype2006,
demeoExtensionBusAsteroid2009,demeoTaxonomicDistributionAsteroids2013,
demeoOlivinedominatedAtypeAsteroids2019,choiTaxonomicClassificationAsteroids2023}. PCA allows to reduce the dimensionality of the data while preserving most of the variance, enabling the identification of clusters corresponding to different taxonomic classes.

Here, we apply a probabilistic, data-driven classification approach based on PCA derived from Gaia DR3 reflectances. Our method is inspired by the results of \citet{bhatSearchingPrimitiveDark2025} and \citet{delboGaiaIRTFAbundance2026} that showed how asteroids of different spectral classes cluster in distinct regions of the principal component space.
Using asteroids with known spectroscopic classes as reference set, we construct multi-dimensional probability density functions via kernel density estimation (KDE) in the principal component space 
(hereafter PC space). 
This allows us to compute, for each Gaia asteroid, the likelihood of belonging to a given spectral complex (C, S, X) or be an end-member (B, V, K, L, A, D) of the Bus-DeMeo taxonomic scheme \citep{demeoExtensionBusAsteroid2009}, and to quantify classification uncertainties.

Our method enables a large-scale spectroscopic classification of Gaia DR3 asteroids based purely on their reflectance properties, with a controlled statistical confidence. The resulting catalogue provides a homogeneous view of the compositional distribution of asteroids throughout the Solar System and establishes a foundation for linking Gaia data to physical surface properties, thermal models' results, and collisional families. A main point of our method is that it is based on Gaia data alone, while it relies on external data to identify where known asteroid classes constitute clusters in principal component space. The classes can be naturally expressed in any chosen taxonomic scheme. Here we used that of Bus-DeMeo 
\citep{demeoExtensionBusAsteroid2009} that is widely used in the community.
In section 2 we present the dataset and preprocessing, in section 3 the methods that we used for the classification, in section 4 the results and control against literature spectra and in section 5 some discussion about the implications of our results.

\section{Dataset and pre-processing}
The Gaia DR3 provides low-resolution spectrophotometric observations in the wavelength range 0.35-1.05 $\mu$m for 60,518 minor bodies \citep{gaiacollaborationGaiaDataRelease2023}. Each asteroid spectrum corresponds to the average of several epoch spectra, which are the individual spectra obtained during each transit of the object across Gaia's spectroscopic focal plane. All mean reflectance spectra are represented using the same set of 16 discrete wavelength bands. Each band is associated with a \emph{reflectance\_spectrum\_flag} (RSF), 
which indicates the reliability of the corresponding reflectance value and its uncertainty: values with RSF=0 are considered validated, RSF=1 suggests possibly poor quality, and RSF=2 marks unreliable data. Previous studies \citep{gaiacollaborationGaiaDataRelease2023,galinierGaiaSearchEarlyformed2023,galinierDiscoveryFirstOlivinedominated2024,galinierAsteroidDifferentiationGaia2024} have shown that Gaia DR3 tends to overestimate reflectance in the reddest spectral bands. Conversely, the two bluest bands frequently suffer from high uncertainties and often have RSF values greater than zero \citep{gaiacollaborationGaiaDataRelease2023,delboGaiaViewPrimitive2023}. Moreover, \citet{tinaut-ruanoAsteroidsReflectanceGaia2023} found that the set of solar analogues adopted in \citet{gaiacollaborationGaiaDataRelease2023} introduces a systematic reddening of spectral slopes at wavelengths shorter than 0.55 µm when compared to results obtained using only Hyades 64, the most Sun-like star in the near-UV. These authors proposed a correction for reflectance at wavelengths shorter than 0.55 µm. In addition, \citet{oszkiewiczSpectralAnalysisBasaltic2023} showed that, when compared to low-resolution spectrophotometric data, the absorption band depths for V type asteroids tend to be underestimated, while band centers are systematically overestimated in Gaia DR3 data, highlighting further limitations in the interpretation of diagnostic spectral parameters from Gaia DR3 reflectances.

To begin with, we assembled a reference set, which is used to construct classes's probability density functions in PC space. We chose to use only asteroids with classifications based on spectroscopy, as these are more reliable than photometry-based ones.
%
%
To do so, we selected all asteroids that have Gaia DR3 reflectance spectra and have classifications from the literature obtained with “Method = SPEC” or “Method = SPEC+PHOT” from the global table of asteroid spectral classes available in the MP3C (\url{https://mp3c.oca.eu/properties-search/}). This ensure that we excluded from the latter table classifications derived solely from broad-band photometry (Method = PHOT). This filtering, resulting in 3,981 distinct objects, ensures that the taxonomic labels are based exclusively on higher-fidelity spectral data, thereby reducing ambiguities associated with photometry-only classifications and improving the reliability of the taxonomic assignments. From this list, we extracted those asteroids that have Gaia DR3 reflectance spectra with Signal to Noise ratio, S/N, $\geq$50, where the S/N is defined as in \citet{gaiacollaborationGaiaDataRelease2023}, yielding a reference set of 2,653 asteroids belonging to the main compositional complexes (C, X, and S), as well as to major end-member groups (D, L, K, A, and V), ensuring broad coverage of the spectral diversity of asteroids. For these reference asteroids, we extracted their “best” spectroscopic class from the \url{https://mp3c.oca.eu/best-values-search/}, which is a single spectral type assigned to each asteroid based on the most reliable spectroscopic classification available in the literature. The methodology for determining the "best" value for each property is described in \citet{dyerDependenceAsteroidRotation2026}.

\subsection{Grouping of Spectral Classes}
To facilitate the analysis of asteroid spectral properties, we grouped individual spectral classes into broader taxonomic categories based on their compositional and spectral similarities (Table~\ref{tab:classMapping}). Specifically, subclasses within the C-complex (such as Cb, Cgh, Ch, Cg) were merged into a single C category, reflecting their similar carbonaceous composition and spectral behavior. The B class remained a standalone category and was not merged into the C group. Likewise, subclasses within the X-complex (Xc, Xe, Xk, Xn) were consolidated into a single X category, capturing their redder than C and mostly featureless spectral characteristics. For the S-complex, multiple subtypes exhibiting silicate-dominated spectra (S, Sq, Sr, Sv, Sa, Sw, Svw, Sqw, Srw and intermediate classes such as Sq/Q) were merged into a single S category, preserving the characteristic olivine–pyroxene features that define this class. Several asteroids, mostly near-Earth ones, have been broadly classified with a complex and not a specific type. Therefore, the C, X, and S-complex classification is included in the respective categories.

Certain ambiguous or transitional classifications (D/T, T/D, C/X, T) were either reassigned to the closest dominant type or excluded from the dataset when their spectral identity could not be reliably determined. For instance, C/X class was not mapped to avoid introducing uncertainty into the analysis. This is because by the C/X classification the authors mean an intermediate spectrum between a C and X complex and could indicate different composition.

This grouping approach allows us to reduce the complexity of the spectral class database  while maintaining the essential compositional distinctions among asteroid types. By mapping the literature-based classes to these broader categories, we ensure consistency in subsequent analyses of physical properties, spectral trends, and compositional correlations across the asteroid population. Table~\ref{tab:classMapping} presents the adopted class mapping.

In the end, the reference set is defined on 2,653 asteroids having Gaia DR3 reflectance spectra with S/N$\geq$50 and a spectral classification from the literature based on spectroscopic data. Their literature classes were mapped as in Tab.~\ref{tab:classMapping}.

{
\renewcommand{\arraystretch}{1.2}
\begin{table}[ht]
\centering
\begin{tabular}{l l}
\hline
\textbf{Original Classes} & \textbf{Mapped Classes} \\
\hline
C-complex, C, Cb, Cgh, Ch, Cg, Caa, Cgx, C0 & C \\
X-complex, X, Xc, Xe, Xk, Xn, Xt, Xk/Xn, E, M, P, EMP, X/E & X \\
S-complex, S, Sq, Sr, Sv, Sa, Sw, Svw, S/Sq, Sq/Q, Sqw, Srw, Srq, Sk & S \\
F, B, B1 & B \\
D, D0, D1, D2, D3, Z, D/T & D \\
L & L \\
K, K1 & K \\
A & A \\
V & V \\
T/D, T & T \\
Q & Q \\
R & R \\
O & O \\
\hline
\end{tabular}
\caption{ Systematic mapping to reduce the complexity of the dataset while maintaining key compositional distinctions. The left column represents the unique classes found in the best class from the MP3C following the method presented in \citet{dyerDependenceAsteroidRotation2026}.}
\label{tab:classMapping}
\end{table}
}

\section{Methods} 
We reduced the dimensionality of the Gaia DR3 reflectance spectra and emphasize its main sources of spectral variance by applying the PCA method to the Gaia DR3 reflectance spectra as it was done by \citet{delboGaiaIRTFAbundance2026} and \citet{bhatSearchingPrimitiveDark2025}. Before the PCA, we did not to apply the correction of \citet{tinaut-ruanoAsteroidsReflectanceGaia2023} as our approach relies on a supervised classification method in which class probabilities are assigned to Gaia DR3 asteroids based on the location of objects with known spectral classes -- the reference set -- from the literature in the PCA components' space. Consequently, the classification is performed entirely using Gaia DR3 data, without comparison of the shapes of Gaia DR3 reflectance spectra against external spectral datasets.
We did not consider the two bluest and the two reddest bands and the values with RSF$>$0. We replaced reflectance values at bands with RSF = 0 by values interpolated from neighboring bands with RSF $>$ 0 using cubic splines.

We applied a PCA with five components. However, it has been shown that if the PCA is performed on those reflectance spectra with good (S/N)$>$50 (i.e. 9,332 objects), more than 95\% of the variance is contained within the first three components \citep{delboGaiaIRTFAbundance2026}. Hence, we decided to carry on the analysis only with three components PC1, PC2, PC3. Having the fit established, all asteroids having reflectance spectra within Gaia DR3 (regardless of their S/N, i.e. 60,518 minor bodies), were then represented as a point in the three-dimensional (PC1, PC2, PC3) space. It has been shown that the spectroscopic C, X, and S complexes as well end-members classes such as B, D, A, and V form distinct clusters in this PC space \citep{delboGaiaIRTFAbundance2026,bhatSearchingPrimitiveDark2025}. In summary, the transformation from the reflectance space to the PC space is determined using 9,332 asteroids with S/N$>$50; this transformation is applied to 60,518 asteroids.

For each spectral class listed in the right column of Table~\ref{tab:classMapping}, we model the distribution of objects of the reference set in PC space using a multivariate KDE, thereby capturing the empirical probability density associated with each class. All Gaia DR3 asteroids can then be classified by evaluating the KDE of each spectral class at the position of a given object in PC space (Eq.~\ref{eq:kde}), multiplying this value by the relative abundance of objects in that class, and normalizing over all classes to obtain the probability of membership in each class. This approach provides a consistent, data-driven, and probabilistic taxonomic classification anchored to high-quality spectroscopic measurements. Details are given in the following subsections.

\subsection{Probabilistic classification via kernel density estimation}
For each spectral class $c$, the KDE estimates the continuous probability density function $f_c\left(x\right)$ of the distribution of asteroids of the reference set in the three-dimensional space defined by $(PC1, PC2, PC3)$. The KDE function is given by Eq.~\ref{eq:kde}:
\begin{equation}
    f_c\left(x\right)=\frac{1}{N_ch^3}\sum_{i=1}^{N_c}K\left(\frac{x-x_i}{h}\right)
    \label{eq:kde}
\end{equation}
where $N_c$ is the number of asteroids of the reference set that have a Gaia DR3 reflectance spectrum with S/N$>$50 in the class $c$, $x_i=\left(PC1_i,PC2_i,PC3_i\right)$ are the coordinates of the asteroid $i$ in the reference set, $h$ is the kernel bandwidth, and $K$ is a Gaussian kernel. Only classes with at least five objects were considered for the KDE fitting. While this threshold is arbitrary, it is a reasonable compromise to ensure that each class has enough samples for a robust KDE estimation. The bandwidth $h$ was optimized using cross-validation on the reference set to balance bias and variance. The number of objects in each class, $N_c$, is given in Table~\ref{tab:refAstNumber_KDE}. Its comparison with Table~\ref{tab:classMapping} shows that the T, Q, R, and O classes were not included in the KDE fitting due to insufficient numbers of objects with high-quality Gaia DR3 spectra. In the end, 2,447 asteroids from the reference set were used for the KDE fitting.

\begin{table}[h]
\centering
\begin{tabular}{lc}
\hline
Class & $N_c$ \\
\hline
S & 887 \\
C & 537 \\
X & 457 \\
D & 183 \\
B & 107 \\
V & 103 \\
A & 80 \\
L & 57 \\
K & 36 \\
\hline
\textbf{Total} & \textbf{2447} \\
\hline
\end{tabular}
\caption{Number of asteroids per spectral class used to fit the KDE models sorted by the number of objects in each class. The number of objects per class is denoted as $N_c$. }
\label{tab:refAstNumber_KDE}
\end{table}

For each unclassified asteroid $j$ with coordinates $x_j$ in PC space, we compute the likelihood $P_c\left(x_j\right)$ of belonging to class $c$ from Eq.~\ref{eq:classP}:
\begin{equation}
    P_c\left(x_j\right)=\frac{f_c\left(x_j\right)}{\sum_{k}f_k\left(x_j\right)}
    \label{eq:classP}
\end{equation}
where at the denominator the sum, $k$, is over all classes. 
The final classification is assigned to the class with the maximum probability by means of Eq.~\ref{eq:classPmax},
\begin{equation}
c^\ast = \arg\max_{c} \, P_c(x_j),
\label{eq:classPmax}
\end{equation}
and the probability itself provides a measure of the classification confidence. This means that for each asteroid, the values of $P_c(x_j)$ across all classes are evaluated and the classes with the highest value and the second-highest value are selected. Their $P$-values, allowing us to quantify the relative likelihood of each taxonomic assignment and to identify cases where the classification is uncertain (e.g., when multiple classes have similar probabilities), are also reported.
This approach naturally handles overlapping class regions in PC space and provides continuous probabilities rather than discrete labels, making it particularly well-suited to large and noisy datasets such as Gaia DR3. The application of this method to the PCA values from Gaia DR3 yields class-probability values for each asteroid. 

\subsection{Validation and performance method}
To assess the performance of our spectral classifier, we used a validation set consisting of asteroids with reliable “known" spectral class from the literature. Using the best-value table of the MP3C, we selected asteroids with classification score $\geq$10 \citep{dyerDependenceAsteroidRotation2026}, totaling to 11,258 asteroids. Of these set we retained those asteroids with Gaia DR3 reflectance S/N$\geq$30 (5,689 objects); we then computed their probability distribution across all classes of Table~\ref{tab:refAstNumber_KDE} and assigned to each asteroid a predicted class corresponding to the class with the probability maximum. While S/N$\geq$30 is a somewhat arbitrary threshold, it represents a reasonable compromise to ensure that the validation set includes a sufficient number of objects with reliable spectral data while still being large enough to provide meaningful statistics for the confusion matrix analysis.
There are 2,114 asteroids in the validation set that are also in the reference set.

We then constructed a confusion matrix (Fig.~\ref{fig:confusion_matrix}) that compares the predicted classes to the reference classes. We normalized the matrix by rows, so that each row represents all objects predicted to belong to a given class and the values indicate the fraction of these predictions that correspond to each true taxonomic type. This “precision-style” normalization allows us to quantify how pure each predicted class is and to identify cases of class overlap or systematic confusion between neighboring taxonomic complexes. To preserve interpretability, the matrix was restricted to nine major classes and reindexed according to a fixed ordering.
The resulting heatmap provides an intuitive representation of classifier performance: diagonal elements quantify the fraction of correct predictions for each class, while off-diagonal elements show specific misclassification pathways (e.g., S-A or C-X mixing). This verification procedure confirms both the strengths and limitations of the KDE-based classifier and provides a transparent diagnostic tool for understanding its behavior across the taxonomic space.

\subsection{Implementation details}
All computations were performed in Python, using the pandas, numpy, scikit-learn, and matplotlib libraries. The PCA was implemented using the PCA class from scikit-learn, while the KDE was implemented using the KernelDensity class with a Gaussian kernel. In addition to the KDE, the code stores the number of training objects per class, which later serves as the class prior probability, meaning that classes represented by more objects are given proportionally higher prior weight.

\section{Results}
\begin{figure}[h!]
\begin{center}
\includegraphics[width=\textwidth]{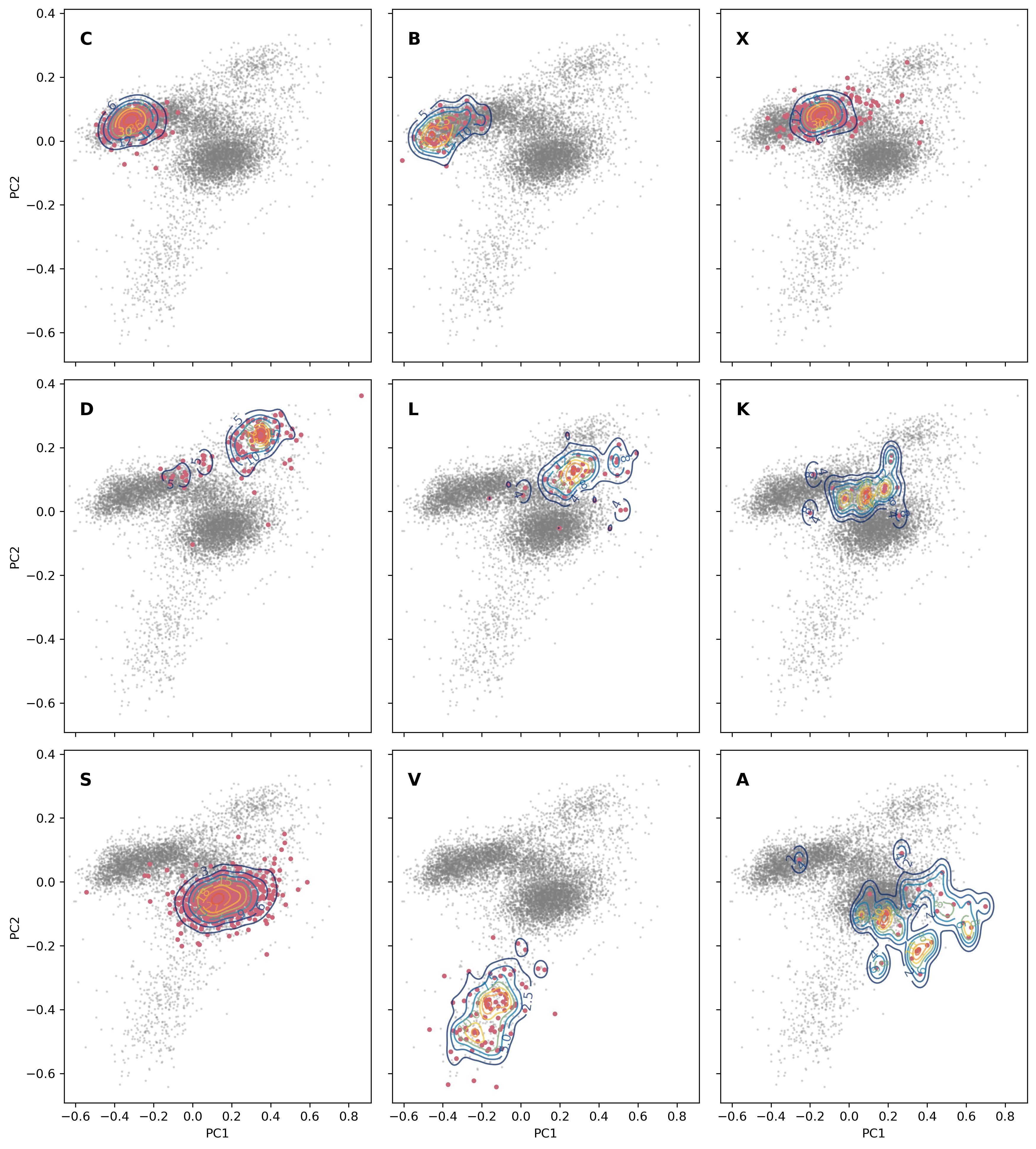}
\end{center}
\caption{Kernel density estimate (KDE) contours in the PC1-PC2 space for different spectral classes indicated by the letters in each plot. Contours represent the estimated density from a Gaussian KDE, providing a smoothed view of the clustering of high-quality (S/N$>$50) reference set asteroids in principal component space for the given spectral class (scatter points).
The background grey points represent the positions in PC1, PC2 space of asteroid with DR3 reflectance spectra with (S/N$>$50).}
\label{fig:KDE_contours}
\end{figure}

Following the methods described above, we begin by defining the optimal value of the bandwidth parameter, $h$, which controls the smoothing of the KDE and thus the level of detail in the reconstructed class distribution. After some trial and error we determined that 0.02$<h<$0.05 gives satisfactory results (Fig.~\ref{fig:KDE_contours} shows the KDEs contours in the (PC1, PC2) space of the nine mapped classes that we used in this work.

\begin{figure}[h!]
\begin{center}
\includegraphics[width=\textwidth]{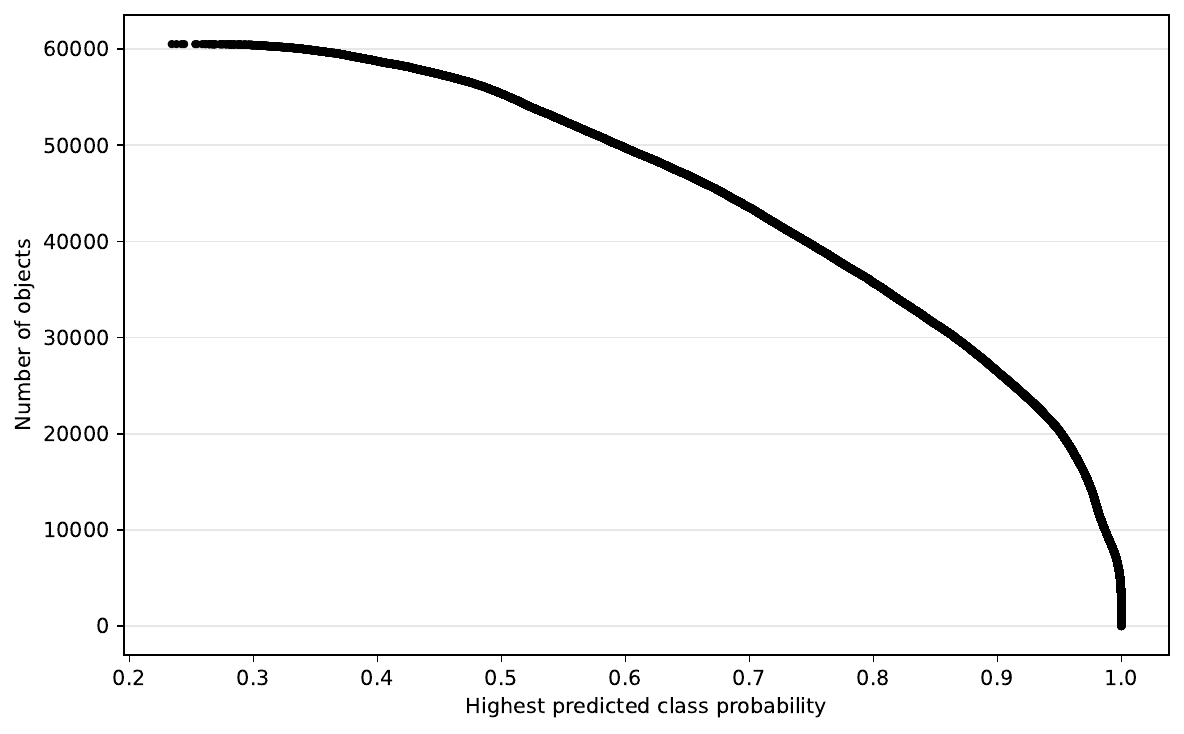}
\end{center}
\caption{Cumulative distribution of the highest predicted class probability for all objects in the sample. Objects are sorted in descending order of their maximum class probability, and the cumulative number of objects is shown as a function of probability. This representation illustrates the confidence level of the classification, highlighting the fraction of objects with highly secure assignments and the gradual transition toward lower-confidence classifications.}
\label{fig:HigestPredictedClassProba}
\end{figure}

Once the KDE models were calculated, we proceeded to classify each asteroid of the 60,518 of the Gaia DR3. The output of the classification is a .csv file listing, for each asteroid, the asteroid number and name, the S/N of the Gaia DR3 reflectance spectrum, the first three principal components (PC1, PC2, PC3), the spectral class with the highest  probability together with its value, and the class with the second-highest probability with its corresponding value. This file constitutes our supervised classification of asteroid spectra based on Gaia DR3 data and is provided as a supplementary file. Figure~\ref{fig:HigestPredictedClassProba} shows the cumulative distribution of the highest predicted class probability.

From visual inspection of the distribution of the asteroids belonging to the different classes in PC1, PC2 space, it is recommended to use classification only for those asteroids with Gaia DR3 reflectance spectra with S/N$\geq$20, i.e., 36,566 asteroids.

\section{Discussion}
To begin with, we present a consideration about our choice of an approach based on the PCA and the KDE. This method provides several advantages in our context. First, the dimensionality reduction offered by PCA allows for a clear and interpretable representation of the spectral data in the PC1-PC2 space, where distinct taxonomic classes naturally cluster \citep{delboGaiaIRTFAbundance2026,bhatSearchingPrimitiveDark2025}, as shown by previous studies. This visualization aspect is important for understanding the structure of the data and for diagnosing classification behavior, whereas ensemble methods such as random forests or neural networks might not offer a straightforward low-dimensional representation and would still require a separate dimensional reduction step for interpretation. Second, the combination of PCA and KDE yields a fully probabilistic, density-based classifier. The probabilities produced by KDE have a clear statistical meaning, reflecting class likelihoods in the reduced feature space. On the other hand, although random forests can output class probabilities, these are not density estimates and do not carry the same interpretative value in terms of probability density or separation between taxonomic clusters. Finally, our dataset is characterized by strongly imbalanced class sizes, with some dominant taxonomic groups (S, X, C) and several much smaller groups (L, D, V). In such cases, density-based classifiers perform robustly. Overall, PCA and KDE provided a good compromise between interpretability, scientific insight, and classification performance for our specific application.

\subsection{Validation of results against known spectral types}
\begin{figure}[h!]
\begin{center}
\includegraphics[width=\textwidth]{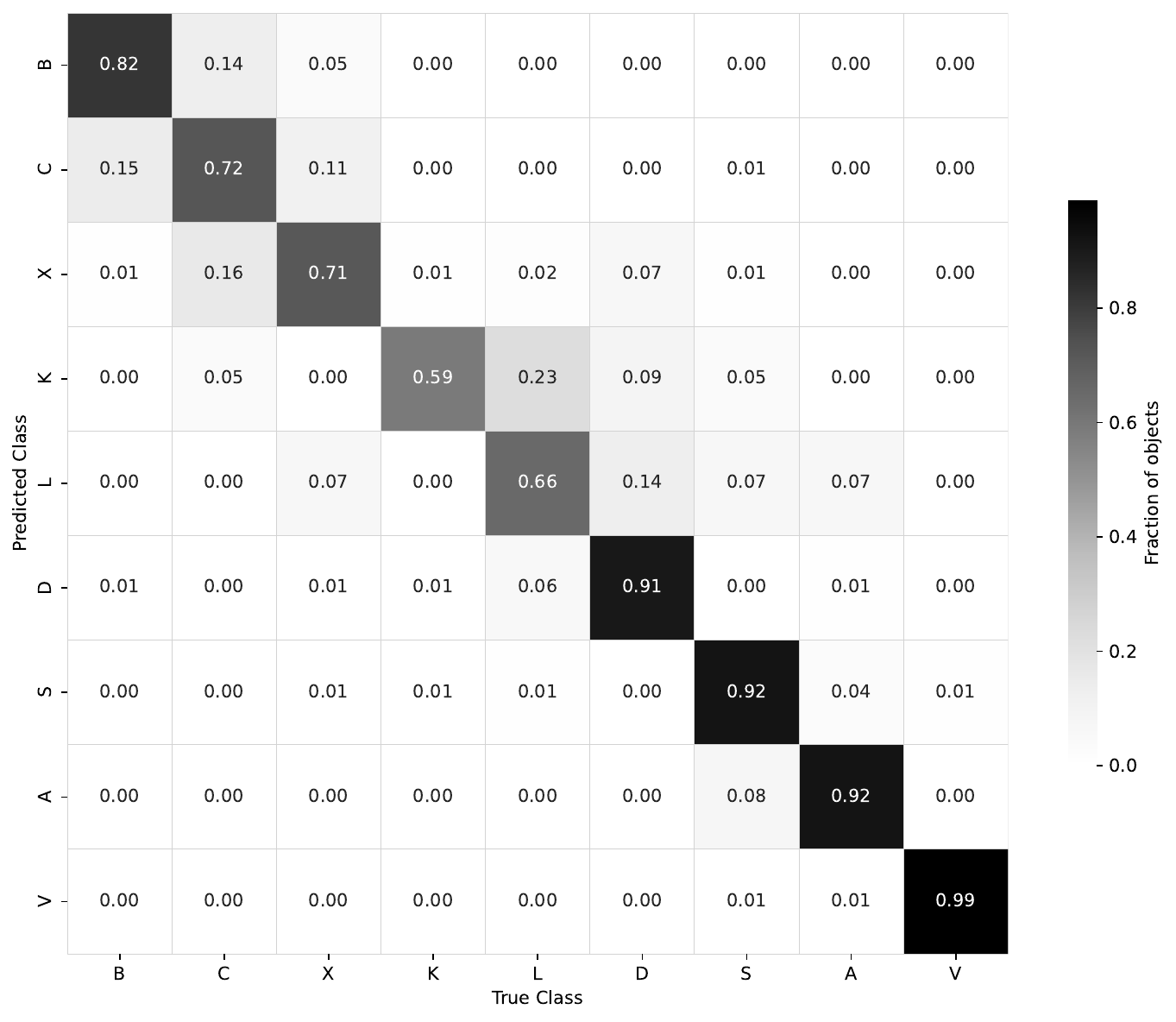}
\end{center}
\caption{Confusion matrix showing the performance of our supervised taxonomic classifier applied to Gaia DR3 asteroid spectra. The matrix displays, for each predicted class (rows), the fraction of classifications that correspond to each true taxonomic type (columns). Values are row-normalized so that each row sums to unity, providing a “precision-style” measure of class purity. Strong diagonal dominance is observed for B, D, S, A, and V types, while expected mixing appears within transitional or spectrally continuous groups such as B-C and K-L-S. This representation highlights both the robustness of the classifier for spectrally distinctive classes and the intrinsic overlaps among intermediate taxonomic types in the Gaia DR3 wavelength range.}
\label{fig:confusion_matrix}
\end{figure}

We applied the validation method described above, resulting in the confusion matrix shown in Fig.~\ref{fig:confusion_matrix}.

The confusion matrix demonstrates that the KDE classifier achieves good performance for several of the major taxonomic groups. The most robust classifications are those of the B, D, A, V types, and S-complex objects, which show strong diagonal dominance. S-complex predictions are correct in more than 92\% of the high-confidence cases, while D type classifications reach 91\%, reflecting the strong spectral distinctiveness of these classes in Gaia DR3 PC space. Similarly, A types and V types—both end-member olivine- and pyroxene-rich classes—are recovered with high purity, at 92\% and 99\%, respectively. These results indicate that the classifier effectively captures the characteristic shapes and slopes of these spectrally extreme types.

The C and B classes also perform well, with C type predictions correctly corresponding to true C types in 72\% of cases, and B type predictions showing a substantial fraction of true B types (82\%). Some mutual contamination is present between B and C types, consistent with their proximity in both reflectance spectral behavior and PC space. This B-C interchange is expected because Gaia DR3 spectra have limited leverage on the spectral slope, which is typically needed to fully separate these classes. There is also some contamination of the C types with X types and vice-versa,  which is expected given the spectral similarity between these two classes.

The X and K/L complex displays the highest degree of mixing. X types are correctly identified in 71\% of the cases, but show leakage toward C types (16\%) and D types (7\%), though the latter is minimal. The K types are also mixed, with correct classifications in 59\% of predictions and significant confusion with L types (23\%) and D types (8\%). L types also retrieved only 66\% of the time, showing partial contamination, mostly with D types (14\%). This behavior is consistent with the known spectral continuity between the K-L-S complex and the fact that Gaia's restricted wavelength range limits the ability to resolve subtle near-infrared band shapes and slopes that distinguish these groups in classical taxonomies. Moreover, the K and L type asteroids are relatively rare, and their classification accuracy appears to be lower. In addition to spectral similarity with neighboring classes, the limited number of samples for these types might also contribute to the reduced performance of the classification.

Overall, the validation confirms the expected strengths and limitations of the PCA+KDE classifier. Classes with distinctive reflectance signatures (A, V, D, and S) are recovered with high purity, while intermediate and spectrally transitional classes (B-C and K-L-S complexes) exhibit the anticipated degrees of overlap. This pattern is fully consistent with previous spectroscopic taxonomic studies and demonstrates that the classifier is performing in line with the intrinsic separability of the classes within the Gaia DR3 spectral domain. 

When comparing with established taxonomic classifications, it is important to note that these classifications are not always 100\% confident in their assignments. Specifically, when we extract "best-values" classes from the MP3C, the Bus-DeMeo taxonomy \citep{demeoExtensionBusAsteroid2009} is favored by construction \citep{dyerDependenceAsteroidRotation2026}. This taxonomy was developed using near-infrared (NIR) spectra. Since Gaia observed only in the visible range, its classifications may differ due to the absence of NIR spectral features that are important in the Bus-DeMeo taxonomy. 

\subsection{Gaia spectroscopic view of major collisional asteroid families}
As an example application (and further validation) of our asteroid spectral classification, we present the taxonomic distribution of major asteroid collisional families.
Asteroid families are the product of catastrophic or cratering collisional events that occurred between small bodies. Collisional fragments that originate from a non-differentiated and homogenous parent body will also appear to have a homogenous distribution of physical properties, such as geometric visible albedos or reflectance spectra. On the other hand, families that originate from partially or fully differentiated planetesimals are expected to show an albedo and spectral diversity. 
The vast majority of asteroid families were identified by using their dynamical characteristics to form clusters in orbital element space \citep{milaniAsteroidFamiliesClassification2014,novakovicAsteroidFamiliesPortal2019,nesvornyIdentificationFamiliesAstIV2015,nesvornyCatalogProperOrbits2024}. Clustering is typically performed in the orbital element regardless of compositionally diagnostic properties, such as albedo and spectral class. This led to the misclassification of some asteroids to be considered family members while in reality are interlopers; objects that happen to be in the orbital neighborhood of the cluster core of the family. The introduction of asteroid physical properties such as albedo and spectra classes into the family identification and characterization studies allow us to “clean” them by eliminating the potentially false membership and most importantly to study the interior of their precursor planetesimals -- in the case of catastrophically disrupted parent bodies. In turn, this enables their link to known meteorites which are the smallest fragments of collisions that survived their impact to Earth \citep[][and references therein]{avdellidouAthorAsteroidFamily2022a, avdellidouDatingSolarSystems2024, brozSourceRegionsCarbonaceous2024, brozYoungAsteroidFamilies2024, marssetMassaliaAsteroidFamily2024}.

\begin{figure}
\begin{center}
\includegraphics[width=0.85\textwidth]{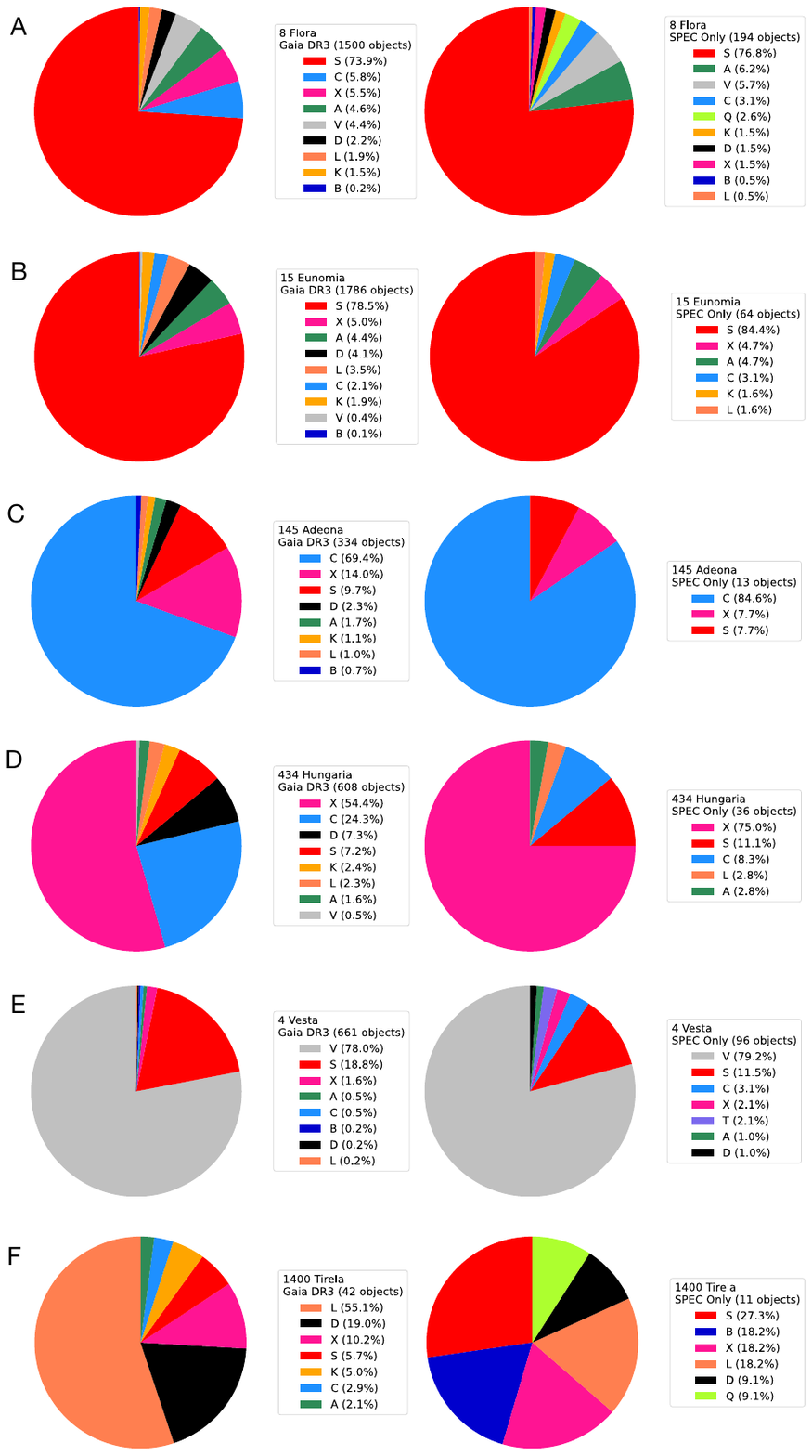}
\end{center}
\caption{Comparison of taxonomic class distributions for members of the asteroid families discussed in this work (see text). Each pie chart shows the fractional contribution of major spectral classes among family members with Gaia DR3 reflectance spectra with S/N$\geq$30. Left column: Gaia DR3 supervised classifications. Right column: Literature classifications based solely on spectroscopic observations (SPEC only), providing the highest-fidelity reference for comparison. Plots labelled A, B, C, D, E, and F correspond to the Flora, Eunomia, Adeona, Hungaria, Vesta and Tirela families, respectively}
\label{fig:family_comparisons}
\end{figure}

Here we selected six asteroid families in order to compare the spectral classes of their members between Gaia DR3 and literature spectra obtained from the ground (Fig.~\ref{fig:family_comparisons}). We did not use spectrophotometric data in order to minimize the probability of misclassification. The selection of the families was done in a way to cover various spectral complexes/classes, different numbers of members and be located in different parts of the main belt.

\emph{Flora family:} Flora is a well-studied inner-main belt S-complex family that has been associated with ordinary chondrite meteorites of type LL \citep{vernazzaCompositionalDifferencesMeteorites2008}. Both Gaia DR3 and ground-based spectra agree that the S-complex dominates the family, with some V type contamination from the nearby Vesta family being noticeable in both samples. Interestingly Gaia DR3 detects similar percentage of olivine-rich A type asteroids than previously reported \citep{delboGaiaIRTFAbundance2026}. The latter study also used Gaia DR3 data and extended spectroscopy into the near-infrared by ground-based NASA's IRTF observations; it was found the abundance of the A type asteroids in the main belt is an order of magnitude larger than what was previously estimated \citep{demeoOlivinedominatedAtypeAsteroids2019}, and that the Flora family appears to have higher abundance of A type compared to the local background population. This is a strong indication that the Flora family originates from a partially differentiated planetesimal.

\emph{Eunomia family:} Eunomia is another well studied central-main belt S-complex family, where the main Gaia DR3 and literature spectra agree well. As Flora, Eunomia has also a significant A type component that has also been reported by \citet{delboGaiaIRTFAbundance2026} to be larger than the local A type background population in the central belt. Another reason that we chose to show the Eunomia family is the significant K type component that appears in Gaia DR3 data. We believe that this could be due to potential misclassification of S-complex asteroids that may appear as K types in Gaia DR3.

\emph{Adeona family:} Adeona is a C-complex family of the central belt. Gaia DR3 shows a significant contamination of X-complex and S-complex objects. This has been first detected and discussed in \citet{bhatSearchingPrimitiveDark2025}, where the authors found a bimodal distribution in the albedo values of the family members. This contamination (along with possibly the much smaller A type contamination) might originate from the nearby S-complex Eunomia family. Indeed, it has been noted that the Adeona family stands out because spectroscopic observations show its members are primarily classified as C and Ch types, whereas the surrounding background population in the same region is largely composed of S type asteroids \citep{hsiehAsteroidFamilyAssociations2018}.

\emph{Hungaria family:} Hungaria is a well studied family in the Hungaria region of the main belt. It belongs to the X-complex while dedicated observations place it specifically in the Xe-class \citep{lucasHungariaAsteroidRegion2017,lucasHungariaAsteroidRegion2019} and has been uniquely linked to the aubrite meteorites, considering also the high geometric visible albedo values ($p_V\gtrsim$ 0.3) of its members. On this, Gaia DR3 agrees somewhat with the ground-based observations. Hungaria is surrounded by an S-complex population \citep{lucasHungariaAsteroidRegion2017} which is evident as family contamination. This contamination part is again probably misclassified as K types in the Gaia DR3 data.

\emph{Vesta family:} Vesta is an asteroid family of the inner main belt that formed after two large basin-forming collisional events. Vesta is a fully differentiated body and the two collisions excavated crust material generating the howdrite-eucrite-diogenite (HED) meteorites \citep{russellDawnVestaTesting2012,mcsweenjr.DawnVestaHED2013}. Both literature and Gaia DR3 observations agree in the prominence of the V types in the family, while the S-complex component is likely due to a contamination from the nearby Flora family. This is reasonable because Vesta is a fully differentiated object with a basaltic crust, and, therefore, an S-complex component that would be associated with chondritic material is exogenous to the collisional population.

\begin{figure}[h!]
\begin{center}
\includegraphics[width=\textwidth]{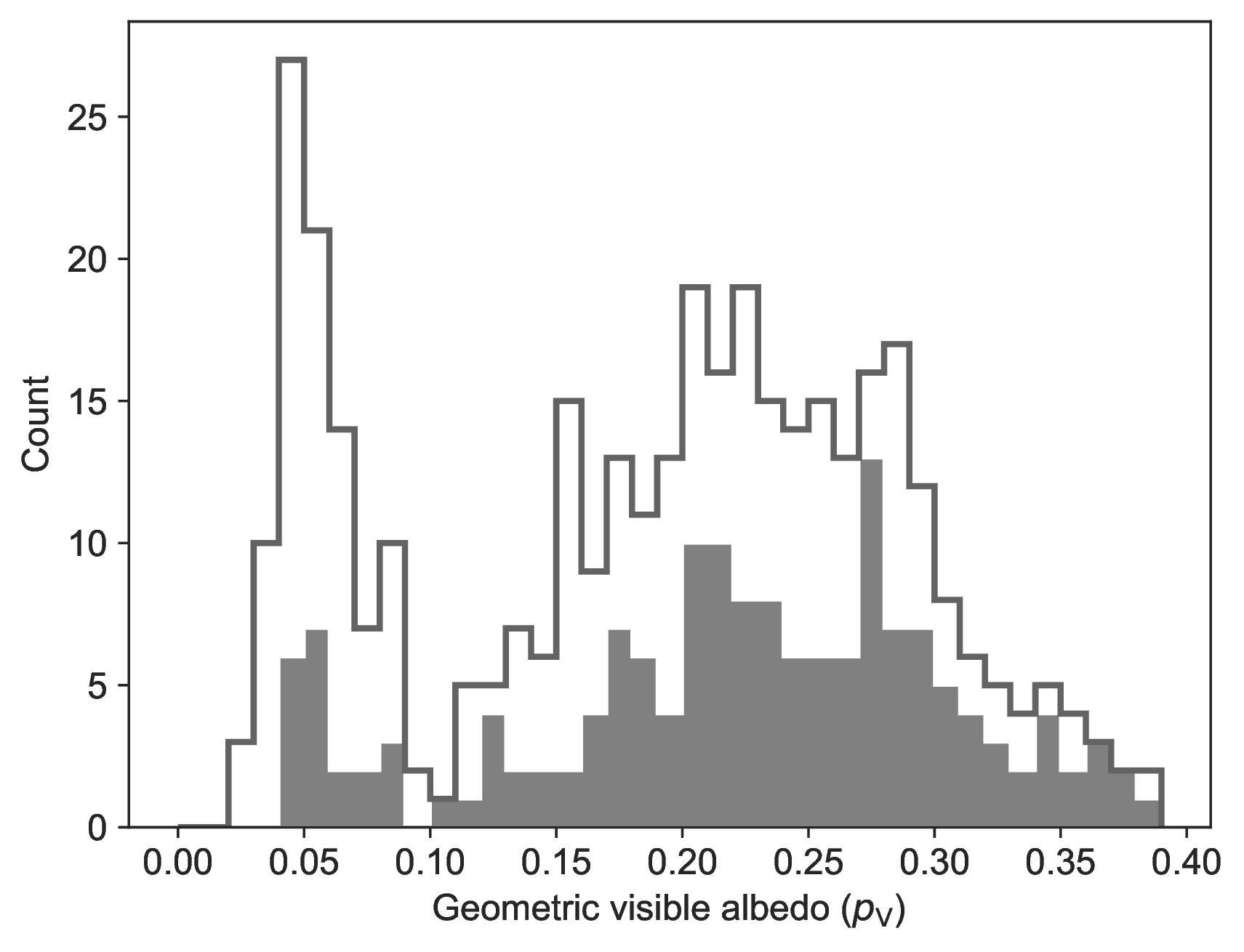}
\end{center}
\caption{Distribution of geometric visible albedo for members of the Tirela asteroid family. Open histograms represent family members with previously known albedo values, while shaded histograms show members for which both albedo and taxonomic classes are derived from Gaia DR3 in this work. High-albedo objects are predominantly L type asteroids, whereas low-albedo objects are mostly D types.}
\label{fig:tirelaAlbedo}
\end{figure}

\emph{Tirela family:} The Tirela family was reported to be mainly composed of Ld type asteroids by \cite{mothe-dinizTirelaUnusualAsteroid2008}, and of L type asteroids by \cite{devogelePhasepolarizationCurveAsteroid2018}. Ld types differ from D types by having a slightly steeper spectral slope shortward of 0.75~$\mu$m and a flatter spectrum long-ward of 0.75~$\mu$m \cite{mothe-dinizTirelaUnusualAsteroid2008}.
\cite{balossiGaiaDR3Asteroid2024} confirmed previous results using Gaia DR3 data and showed that several interlopers — mostly C- and X type asteroids - are present in the HCM definition of the Tirela family. Our classification of Gaia DR3 reflectance spectra reveals that the Tirela family is dominated by L type asteroids ($\sim$55\%), followed by a D type component (19\%). This result is consistent with the albedo distribution of the family as a whole (see Fig.~\ref{fig:tirelaAlbedo}), which shows a prominent peak at a geometric visible albedo of $\sim$0.21, characteristic of L types, and a secondary peak at $\sim$0.05, likely related to C- and D type asteroids.
The source of the albedo values is the MP3C database 
\citep{delboMinorPlanetPhysical2022}, best values table downloaded on December 1, 2025. This table compiles values from various sources including WISE, AKARI, IRAS, and Spitzer, and it is derived as detailed in \cite{dyerDependenceAsteroidRotation2026}. On the other hand, the taxonomic classification of the Tirela family inferred from the 11 asteroids selected, using the method described above, to form our control sample is different (Fig.~\ref{fig:family_comparisons}F, right panel) than the one obtained from our classification (Fig.~\ref{fig:family_comparisons}F, left panel). This discrepancy is likely due to the small number of control objects and to the fact that this sample may be dominated by interlopers. When considering Tirela family members with taxonomic classifications derived from spectroscopy — which constitutes a larger sample than the control sample — we find that most family members are of L or Ld type (Table~\ref{tab:tirelaSpec}). This highlights the importance of using large samples of family members to obtain a reliable view of their compositional distribution.

\begin{table}
\begin{tabular}{r c c c c l}
\toprule
Number & Class$_1$ & Class$_2$ & Taxonomy & Wavelength & Ref. \\
\midrule
3667 & F &  & Tholen & VIS  & \cite{tatsumiUltravioletVisibleSpectroscopy2022} \\
1400 & D &  & Bus & VIS & \cite{lazzaroSOSVisibleSpectroscopic2004} \\
1400 & D &  & Tholen & VIS  & \cite{lazzaroSOSVisibleSpectroscopic2004}  \\
1400 & L & M & Mahlke & VIS  & \cite{mahlkeAsteroidTaxonomyCluster2022} \\
1400 & Ld &  & Bus & VISNIR  & \cite{mothe-dinizTirelaUnusualAsteroid2008} \\
8250 & M & L & Mahlke & VISNIR  & \cite{mahlkeAsteroidTaxonomyCluster2022} \\
9222 & B &  & Bus & VIS  & \cite{mothe-dinizTirelaUnusualAsteroid2008} \\
13150 & Ld &  & Bus & VIS  & \cite{mothe-dinizTirelaUnusualAsteroid2008} \\
19369 & M &  & Mahlke & VISNIR  & \cite{mahlkeAsteroidTaxonomyCluster2022} \\
15552 & L & S & Mahlke & VISNIR  & \cite{mahlkeAsteroidTaxonomyCluster2022} \\
26219 & Q & K & Mahlke & VISNIR  & \cite{mahlkeAsteroidTaxonomyCluster2022} \\
31704 & Ld & D & Bus & VIS  & \cite{mothe-dinizTirelaUnusualAsteroid2008} \\
43343 & Ld &  & Bus & VIS  & \cite{mothe-dinizTirelaUnusualAsteroid2008} \\
44443 & S-complex &  & Bus & VIS  & \cite{mothe-dinizTirelaUnusualAsteroid2008} \\
60157 & Ld &  & Bus & VIS  & \cite{mothe-dinizTirelaUnusualAsteroid2008} \\
60378 & S &  & Bus-DeMeo & NIR  & \cite{humesDistributionHighlyRedsloped2024} \\
68147 & Ld & L & Bus & VIS  & \cite{mothe-dinizTirelaUnusualAsteroid2008} \\
80052 & S &  & Bus-DeMeo & NIR  & \cite{humesDistributionHighlyRedsloped2024} \\
87812 & X &  & Bus-DeMeo & NIR  & \cite{demeoOlivinedominatedAtypeAsteroids2019} \\
84829 & Ld &  & Bus & VIS  & \cite{mothe-dinizTirelaUnusualAsteroid2008} \\
106063 & L &  & Bus-DeMeo & VIS  & \cite{humesDistributionHighlyRedsloped2024} \\
\bottomrule
\end{tabular}
\caption{Spectral classification of Tirela family members from the literature. Class$_1$ and Class$_2$ represent the first and second class assigned to the asteroid in the literature. The taxonomy column indicates the taxonomic system used for the classification. The wavelength column indicates the wavelength range of the spectra used for the classification. Classification based on spectrophotometric data are not included in this table.}
\label{tab:tirelaSpec}
\end{table}

\section{Conclusions}
We have developed and applied a supervised probabilistic framework to classify asteroid reflectance spectra from Gaia DR3, using PCA to compress the spectral information and multivariate KDEs to model the distribution of known taxonomic classes in principal-component space. By relying exclusively on high-quality spectroscopic classifications for the training set, our method provides a homogeneous, data-driven taxonomy that is fully anchored to physical spectral measurements and assigns quantitative posterior probabilities to every asteroid.

Our validation analysis demonstrates that Gaia DR3 data contain sufficient spectral information to reliably separate the major taxonomic complexes. S, D, C, and V types are recovered with good purity, and A types are also robustly identified despite their relative rarity. Mixing between B and C types, as well as between the K–L–S and X complexes, reflects genuine spectral continuity and the limitations imposed by Gaia's restricted wavelength range. These behaviors are consistent with prior spectroscopic surveys and confirm the reliability and interpretability of our probabilistic approach.

The application of our classification to major collisional asteroid families shows strong agreement with published ground-based spectra and highlights new compositional insights enabled by Gaia. In particular, we confirm the enhanced abundance of olivine-rich A type material in the Flora and Eunomia families, supporting the scenario of partially differentiated parent bodies. We also reveal a predominantly L type composition for the Tirela family and identify contamination patterns in Adeona and Hungaria that reflect their dynamical environments.

The catalogue produced in this work—containing posterior probabilities for nine mapped taxonomic classes for all Gaia DR3 asteroids—constitutes an internally consistent spectroscopic classification of these small bodies. It provides a critical bridge between Gaia DR3 spectrophotometry and traditional visible-near-infrared taxonomy, and represents a resource for future studies of asteroid families, compositional gradients, and links between asteroids and meteorites. Forthcoming Gaia data releases will further refine this framework, enabling an increasingly complete and precise spectroscopic mapping of the Solar System's small-body populations.

Future improvements and extensions of the proposed method could proceed along several directions. First, the availability of higher-quality or broader-wavelength spectral data would help reduce degeneracies among taxonomic classes and improve classification robustness. Second, incorporating additional independent constraints, such as thermal-infrared–derived albedos or sizes when available, would allow improved calibration and validation of inferred physical parameters. In addition, Gaia DR4 is expected to provide reflectance spectra together with an unsupervised taxonomic classification. In this context, the supervised approach presented here could play an important role in the validation and cross-comparison of taxonomic classifications derived using different methodologies. Finally, the framework is scalable and could be readily adapted to future large asteroid datasets, provided that suitable training and validation samples are available.

\section*{Conflict of Interest Statement}
The authors declare that the research was conducted in the absence of any commercial or financial relationships that could be construed as a potential conflict of interest.

\section*{Author Contributions}
MD: Conceptualization, data curation, formal analysis, investigation, methodology, project administration, coding supervision, and writing of the original draft and review; TJD: Data curation, visualization, writing and coding; UB: Visualization, writing, and coding; CA: Visualization, Validation, Writing of the original draft, Writing of the review; LG: Conceptualization, data curation, writing; AM: Visualization, writing and coding. All authors contributed to the article, its discussion and approved the submitted and the revised version.

\section*{Funding}
MD is Leverhulme Visiting Professor at the University of Leicester with
financial support from the Leverhulme Trust (UK). MD acknowledges support from the Centre National d'Études Spatiales (CNES) of France. TJD acknowledges financial support from French Space Agency CNES and the Université Côte d'Azur (UniCA). UB acknowledges funding from an STFC PhD studentship. UB thanks the LSST-DA Data Science Fellowship Program, which is funded by LSST-DA, the Brinson Foundation, the WoodNext Foundation, and the Research Corporation for Science Advancement Foundation; his participation in the program has benefited this work. 

\section*{Acknowledgments}
We thank the anonymous reviewers for their constructive comments that helped to improve the manuscript. This work has made use of data from the European Space Agency (ESA) mission Gaia (\url{https://www.cosmos.esa.int/gaia}), processed by the Gaia Data Processing and Analysis Consortium (DPAC, \url{https://www.cosmos.esa.int/web/gaia/dpac/consortium}). Funding for the DPAC has been provided by national institutions, in particular the institutions participating in the Gaia Multilateral Agreement. This work is based on data provided by the Minor Planet Physical Properties Catalogue (MP3C; \url{https://mp3c.oca.eu}) of the Observatoire de la Côte d’Azur.

\section*{Supplemental Data}
 \href{http://home.frontiersin.org/about/author-guidelines#SupplementaryMaterial}{Supplementary Material} should be uploaded separately on submission, if there are Supplementary Figures, please include the caption in the same file as the figure. LaTeX Supplementary Material templates can be found in the Frontiers LaTeX folder.

\section*{Data Availability Statement}
The datasets analyzed for this study can be found in the archive website \url{https://archives.esac.esa.int/gaia} and from the \url{mp3c.oca.eu}.

\bibliographystyle{Frontiers-Harvard} 
\bibliography{test,zotero}

\clearpage


\clearpage


\end{document}